# Single-Shot High-Energy Muon and Particle Radiography with a Multi-GeV Laser-Wakefield-Accelerator-Driven Source

Kaixin Zhu[1], G. Jackson Williams[2], Roberto Versaci[3], Ela Rockafellow[4], Anna Cimmino[3], Jaron E. Shrock[4], Reed Hollinger[5], Gabriele M. Grittani[3], Jiří Šišma[3,6], Ari Sloss[4], Nikola Durand[1], Nischal Tripathi[4] Jay Jablonski[1], James King[5], Gerardo Palma[5], Zach Rautio[1], Bryan Sullivan[5], Shoujun Wang[5], Frederica Sorkin[4], Sina Zahedpour[5], Ping Zhang[5], Bo Miao[4], Andrew Yandow[2], Mayank Gupta[4], Scott W. Hancock[4], Scott C. Wilks[2], Vincent Tang[2], David Warner[1], Jorge J. Rocca[1,5], John Harton[1], Howard M. Milchberg[4], and Brendan A. Reagan[2,5]

[1]Department of Physics, Colorado State University, Fort Collins, CO, 80523, USA

[2]Lawrence Livermore National Laboratory, Livermore CA 94550, USA

[3]ELI Beamlines Facility, The Extreme Light Infrastructure ERIC, Za Radnicı 835, Dolnı Brezany 25241, Czech Republic

[4]Institute for Research in Electronics and Applied Physics, University of Maryland, College Park, Maryland 20742, USA

[5]Department of Electrical and Computer Engineering, Colorado State University, Fort Collins, CO, 80523, USA

[6]Faculty of Nuclear Sciences and Physical Engineering, Czech Technical University in Prague, Brehova 7, 11519 Prague, Czech Republic

We report the first demonstration of single-shot particle radiography using a 1-10 GeV laser-wakefield-generated beam of muons, pions, and neutrons. The test objects were imaged ~15 m from the beam source, through dense lead shielding followed by the walls of a building and a truck. The muon content of the beam was directly confirmed using large volume scintillator-based detectors, which recorded particle decay events with timing delays consistent with the muon lifetime. Simulations confirm that the high energy component of the beam transmitted through the test object is nearly entirely composed of muons, directly showing their highly penetrative nature, with a single-shot fluence equivalent to >8 hours of integration of cosmic ray muons near the horizon. Our work establishes single-shot high-energy particle radiography with a laser-wakefield-accelerator-driven source.

## I. INTRODUCTION

Muons are unstable, fundamental particles with charge equal to electrons and positrons but ~207 times more massive. High-energy muons (~1 GeV and above) are highly penetrating through extended lengths of dense material and can enable radiography of natural and man-made structures that cannot be probed by other forms of radiation. Interest in imaging methods using muons, known as "muography", dates back to Alvarez's proposal to employ cosmic-ray generated muons to nondestructively probe the interior of the Egyptian pyramids [1]. This proposed method culminated in the 2017 discovery of a previously unknown, hidden chamber in the Great Pyramid when a multinational team used an array of detectors to detect muons passing through the structure [2]. Multidisciplinary applications employing muon radiography and tomography have been proposed or demonstrated including probing the interior of volcanos [3,4], cargo container screening and nuclear material detection [5–7], mineral exploration [8], and nuclear reactor imaging [9,10]. However, the applications demonstrated to date have relied upon muons naturally generated by the interaction of cosmic rays with atoms in the upper atmosphere, which have the relatively low flux of ~1 muon/cm$^2$/min at sea level near the zenith angle, decreasing to ~0.1/cm$^2$/min near the horizon [11]. This low flux has limited applications to date. For radiography applications that have been demonstrated, such as the pyramid results, it has required months of data acquisition time.

Artificial muon beam sources currently rely on conventional radiofrequency (RF) accelerators [12–18]. In these sources, high-energy proton beams impinge on a target, producing muons primarily through the decay of secondary pions generated in hadronic interactions. High-energy electron beams irradiating a target can also be used to generate muons through two principal channels: direct muon-pair production via the Bethe–Heitler (BH) process [19,20] and pion production followed by decay [21–25]. In the BH process, bremsstrahlung photons generated in the electromagnetic shower convert into lepton–antilepton pairs and retain the forward directionality of the electron. The threshold for muon-pair production is approximately 211 MeV and the cross section of the process is smaller than electron–positron pair production by the square of the mass ratio, $(m_e/m_\mu)^2 \approx 2.3 \times 10^{-5}$. Alternatively, photons in the shower may produce pions via a photo-nuclear interaction, which subsequently decay into muons. Because photo-pion production has a significantly larger cross section than direct BH muon-pair production at GeV-scale energies, muon yields are generally dominated by pion decay. Efficient muon production requires electron energies from several hundred MeV to multi-GeV, depending on the production channel and desired yield. Conventional accelerators capable of producing such beams typically require facilities ranging from hundreds of meters to kilometers in length.

In contrast, laser wakefield accelerators (LWFAs) can sustain accelerating gradients approaching 100 GeV/m [26], ~1000 times greater than the maximum currently achievable with conventional accelerators, making possible the realization of compact and transportable sources of high flux muon beams for radiographic and other applications [27,28]. Recently, there have been several demonstrations using petawatt-class lasers to accelerate electrons to the 10 GeV level [29–31]. While multi-GeV electrons have been generated with relativistic self-guiding [32,33], preformed plasma waveguides [34] allow for significantly longer acceleration lengths, improved laser-to-electron energy efficiency, and control of optical mode and injected charge [35–39]. The use of all-optical techniques for tunable waveguide generation [36–39] has been instrumental in producing multi-GeV electron bunches with sub-Petawatt laser pulses [30,31,35,38]. Production of both positrons and pions using laser-driven electron beams has previously been demonstrated [25,40–43]. Recently developed meter-scale gas jets [44] offer the potential to accelerate electrons to ~100 GeV in a single [45] or multiple [46] acceleration stages, enabling the realization of a compact, bright muon source [45]. The availability of sufficiently energetic LWFA electron bunches has led to the development of laser-based muon sources at the Texas Petawatt la-ser [47], the BELLA laser [48], the Shanghai Ultra-Intensive Ultra-Short Laser Experiment Facility (SULF) [49], and at the Extreme Light Infrastructure Nuclear Physics (ELI-NP) facility [50]. Additionally, a LWFA-driven source of neutrons was recently demonstrated at Institut National de la Recherche Scientifique (INRS) [51].

Here we demonstrate, for the first time to our knowledge, single-shot laser wakefield-driven high energy particle radiography, performed by imaging dense materials located 15 meters from a laser wakefield electron accelerator. The mixed muon, pion, and neutron beam used in the imaging experiment was produced using 1-10 GeV LWFA-generated electron beams interacting with a tungsten target. The production of muons was conclusively verified through measurement of their characteristic lifetime, along with characterization of the accompanying Bethe–Heitler positron production. Because of the nuclear interaction length of tens of cm for charged pions in dense material combined with their lifetime of tens of ns in the rest frame [52], together with the neutron flux fall-off, a LWFA-driven multiparticle source could be muon-dominated at object thicknesses greater than ~1 m or at tens of meters of distance propagation.

## II. EXPERIMENTAL SETUP

### A. Laser Wakefield Acceleration

Experiments were performed at the ALEPH Laser Center at Colorado State University [53] during the summer and spring of 2024 with the experimental configuration shown in Fig. 1. Multi-GeV electron beams were generated using a 30 cm laser wakefield accelerator operated in an optically formed plasma waveguide as described in Refs. [30,38]. Detailed descriptions of the waveguide formation and accelerator physics can be found in Refs. [34–39,44,54,55] and are only briefly summarized here.

For the present experiments, a 30 cm plasma channel was formed in a 1%/99% N2/H2 gas sheet produced by a supersonic slit nozzle [44]. The waveguide mode size was matched to the drive-laser focal spot ($w0 \approx 30$ μm) by controlling the energy deposited into the plasma by a J_0 Bessel beam, and the delay between the J_0 and drive pulses, enabling efficient coupling of the drive pulse into the channel. Approximately 1 J of laser energy with ~100fs FWHM duration focused with a transmissive axicon was used to generate the plasma/neutral prepared index structure for waveguide formation by the self-waveguiding drive pulse [37]. A logarithmic diffractive axicon was used for the imaging results discussed below, which demonstrated improved electron beam stability by providing robustness to drive pulse misalignment [56]. The self-guided LWFA drive pulse contained up to 20 J in ~40 fs FWHM duration and was focused by an f/25 off-axis parabola into the entrance of the channel.

The use of self-waveguiding LWFA enabled the production of multi-GeV electron beams with charge ranging from tens to hundreds of pC [38]. In recent experiments using this and similar accelerator designs, electron energies of 10 GeV have been demonstrated [30,31]. The electron beams produced in the present campaign were characterized using the diagnostics described below.

### B. Electron Beam Characterization

Accelerated electrons were characterized using a modular permanent magnet spectrometer instrumented with a collimating slit (~0.2 mrad acceptance) and Lanex scintillating screens positioned before the collimator and at the exit plane of the magnetic field [57]. The entrance screens provided measurements of the electron beam profile and pointing stability as well as the coupling efficiency through the slit, while the exit screen enabled reconstruction of the energy spectrum after magnetic dispersion. The collimating slit consisted of 100-mm-thick, fine ground tungsten blocks separated by 0.5 mm stainless steel shims. Since typical multi-GeV electron beams (divergence ~ 1 mrad) overfilled the narrow aperture (~0.2 mrad), only a small fraction of total beam electrons were sampled by the collimator and dispersed in the magnetic field, with the rest terminating in the bulk. This enabled use of the collimator as a converter for secondary radiation generation and the simultaneous characterization of electron beam properties. Use of scintillating screens at the collimator entrance and upstream enabled estimates of the total beam charge on a given shot [30,38].

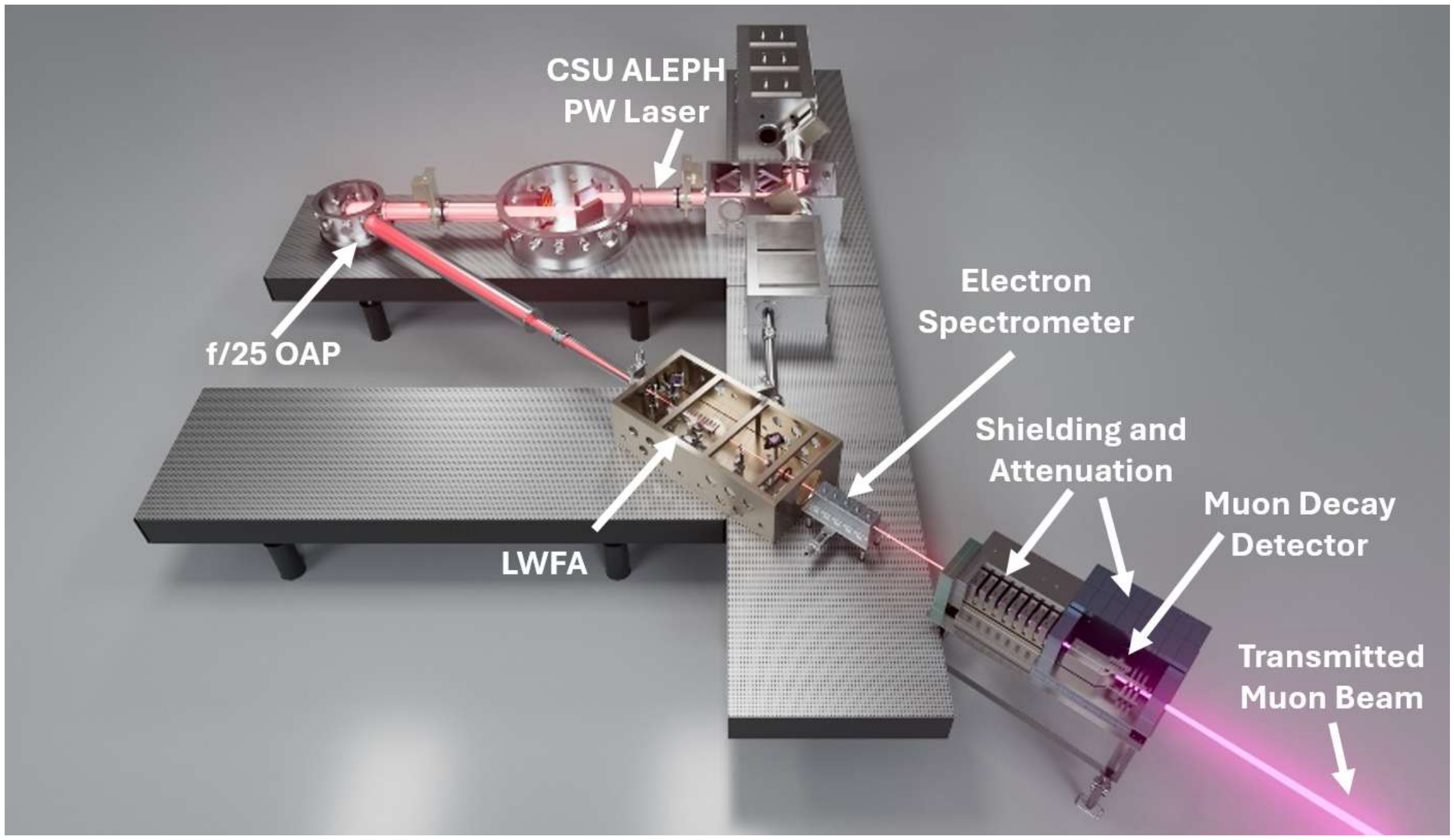

FIG 1. Conceptual rendering of the experimental setup employed for the generation and detection of laser-driven muons and electron positron pairs. λ=800nm laser pulses from the CSU ALEPH laser system were split and sent to two compressors. Approximately 1 J, sub-100 fs FWHM pulses were focused using a diffractive axicon into a 30 cm long nitrogen/hydrogen gas jet to produce the assisted self-guiding structure described in the text. The bulk of the laser energy, up to 20 J, ~40 fs FWHM duration were focused with a f/25 off-axis parabola into the entrance of the gas jet. The accelerated electron spectrum and charge were measured using a magnetic spectrometer with a lanex plate imaged onto a CCD detector. Following the magnetic spectrometer, the majority of the electrons are stopped by a beam dump consisting of lead and high density polyethylene. A shielded detector consisting of a large volume EJ-200 scintillator plastic coupled to a 125 mm diameter photomultiplier tube (PMT) was used for the muon detection via muon decay described below.

Representative electron spectra measured during the experiment are shown in Fig. 2. The data comprise more than 1150 electron beams recorded on a single day and illustrate the range of electron-beam conditions employed in the present study. Since the experimental platform enabled simultaneous measurement of electron beam properties and secondary radiation, key accelerator parameters such as drive pulse length, plasma density, and injector gas were varied substantially throughout each day of operation, sometimes to regions of parameter space where beam generation was suppressed. Additionally, shot-to-shot variation in laser pointing could occasionally result in poor channel coupling, and thus fail to produce multi-GeV electron beams [35,38]. For the representative data in Fig. 2, $\gtrsim$ 70% of laser shots throughout the day produced multi-GeV electron beams, while $\gtrsim$ 80% of laser shots under optimal laser-plasma conditions produced beams. As can be seen from this figure, many beams featured electron energies in excess of 5 GeV, and the mean total charge over this series was estimated to be greater than 100 pC per shot. To maximize the accelerated charge, ionization injection was employed by doping the full 30 cm accelerator length [30,35]. The interaction of the intercepted multi-GeV electrons with the tungsten resulted in an intense secondary particle cascade containing photons, electrons, positrons, pions, neutrons, and muons, which formed the basis of the measurements described below.

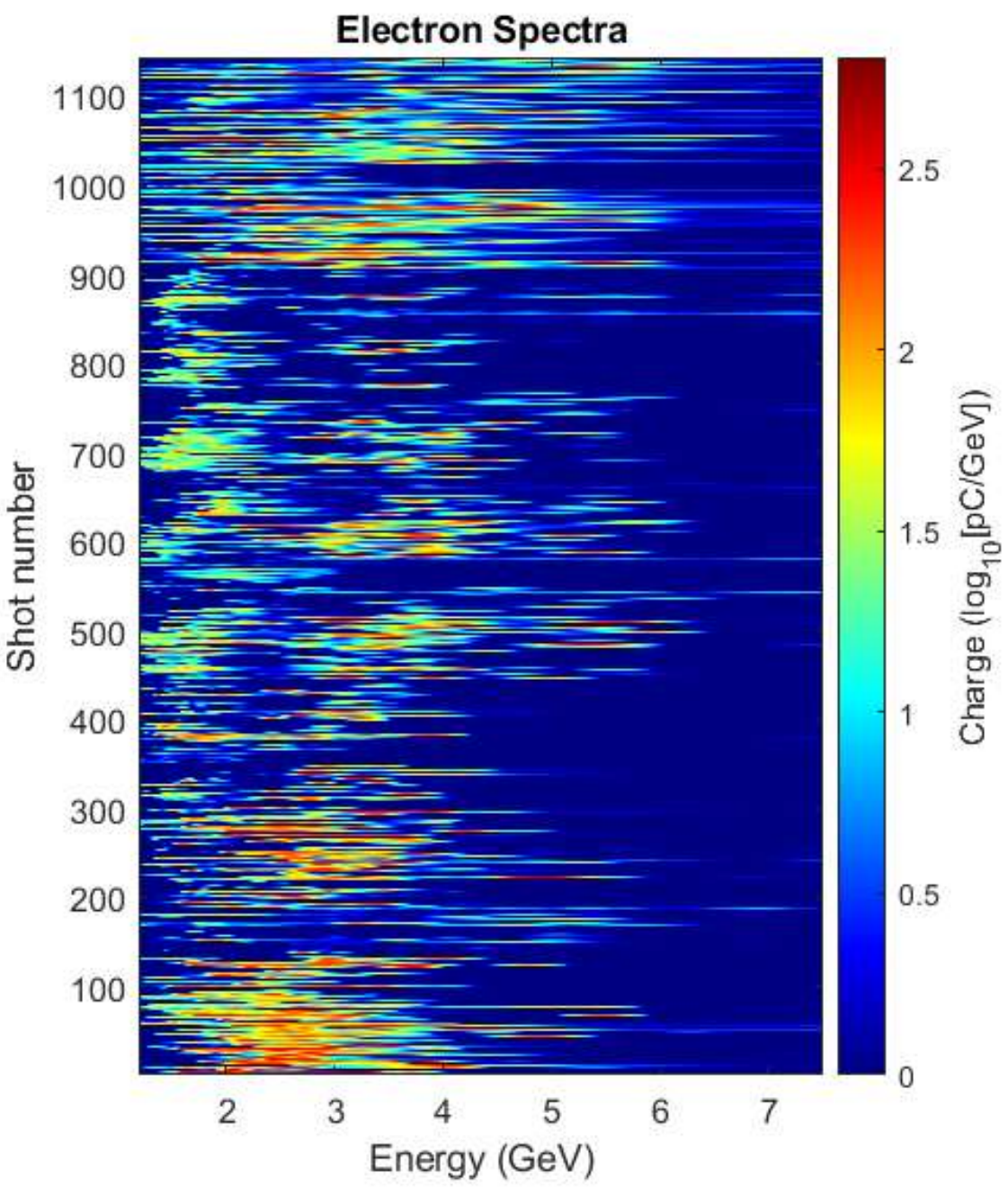

FIG 2. Representative set of 1150 measured consecutive electron spectra with energies ranging between 1-8 GeV plotted on a log color scale for clarity.

### C. Positron Detection System

Positron production was measured using a dedicated magnetic spectrometer configuration, illustrated in Fig. 3(a). For these measurements, the four module GeV-class electron spectrometer was replaced by a single magnet module, 100 mm long, with a mean magnetic flux density of ~1 T on the center axis. A tungsten input slit assembly consisted of vertical and horizontal crossed slits, 25 mm thick each, with 1.3 mm shims. Tungsten converter targets of varying thickness were inserted into the LWFA electron beam upstream of the magnet in order to generate positrons through Bremsstrahlung and Bethe–Heitler pair production.

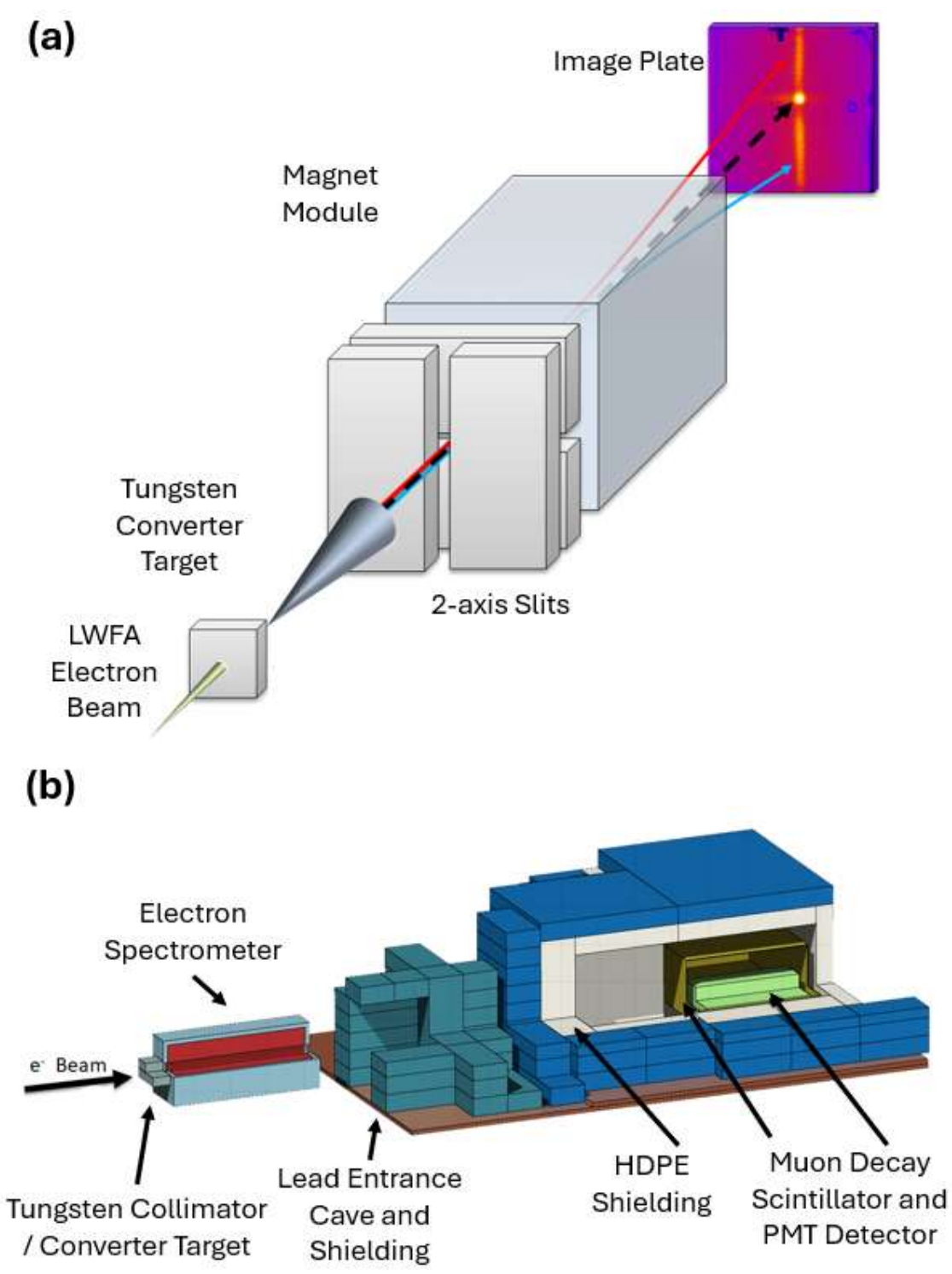

FIG 3. (a) Setup for the generation and detection of laser-driven electron-positron pairs. (b) LWFA-driven muon detection setup.

Electrons and positrons emerging from the converter were deflected in opposite directions by the magnetic field and projected onto phosphor image plates, shown in Section 3. These measurements provide an experimental benchmark of electromagnetic shower development and Bethe–Heitler pair production and validated Monte Carlo simulations of expected muon yields.

### D. Muon Detection System

Muon measurements were performed using the setup shown in Fig. 3(b). Secondary particles generated at the tungsten slit assembly propagated downstream of the spectrometer and beam dump. Electrons transmitted through the spectrometer were deflected into a shielding assembly consisting of lead and plastic. Monte Carlo simulations using Geant4 [58] and FLUKA [59,60] were employed to optimize the detector geometry, shielding configuration, and radiation-safety requirements. The final shielding assembly consisted of 300 mm of lead and 100 mm of high-density polyethylene (HDPE), substantially reducing the flux of photons and neutrons reaching the detector region.

The muon detector was positioned approximately 3 m downstream of the spectrometer and consisted of a 75 mm × 150 mm × 300 mm volume of EJ-200 plastic scintillator optically coupled to a 125 mm diameter Hamamatsu R1250 photomultiplier tube (PMT) whose scintillation light output was collected using a 3 GHz LeCroy 7300A oscilloscope and a 20MHz Pico Technology 4444 oscilloscope. The detector was enclosed in a light-tight housing surrounded by 100 mm of lead and at least 50 mm of HDPE on all sides. Muons entering the scintillator may lose sufficient energy to stop within the detector volume and subsequently decay, producing a characteristic delayed signal that can be identified through its temporal signature. The detector response was calibrated in situ using cosmic-ray muons recorded with the laser off, which resulted in a measured decay time of $2.12 \pm 0.49$ μs from 136 cosmic ray candidates, in agreement with the Particle Data Group (PDG) value of $2.1969811 \pm 0.0000022$ μs [52]. Additional details on the muon detectors are presented in Appendix A.

## III. GENERATION OF ELECTRON-POSITRON PAIRS AND LASER-DRIVEN MUONS

The results of LWFA electron beam driven electron-positron pair production are shown in Fig. 4(a-d), which shows the measured electron and positron spectra integrated over 20 laser shots. For the data set below, ~6-8 of these laser shots resulted in GeV-level electron beams hitting a tungsten converter target with thickness ranging from 2 – 20 mm. Positrons, electrons, and x-rays were measured with phosphor image plates and dispersed using an additional magnet module (see Fig 3(a)). As can be seen from the plot of Fig 4(e), we observe the onset of significant positron-electron pairs at the thinnest targets to a regime that creates a quasi-charge-neutral pair beam in the thickest. This data is in general agreement with previous studies performed with lower energy electron beams [61]. As the target thickness increases, primary electrons lose energy and scatter into high angles away from the spectrometer and more positron-electron pairs are created. At >10 mm, a significant population of the pairs, with lower energy than the primary electrons, do not have enough kinetic energy to escape from the converter target. Positron-electron pairs are produced through an analogous process to BH muon pair production, but with $>10^5$ higher cross sections and are more readily characterized and detected compared to muons. Monitoring the generated positron flux and spectrum can provide a surrogate method for estimating the generated muon flux.

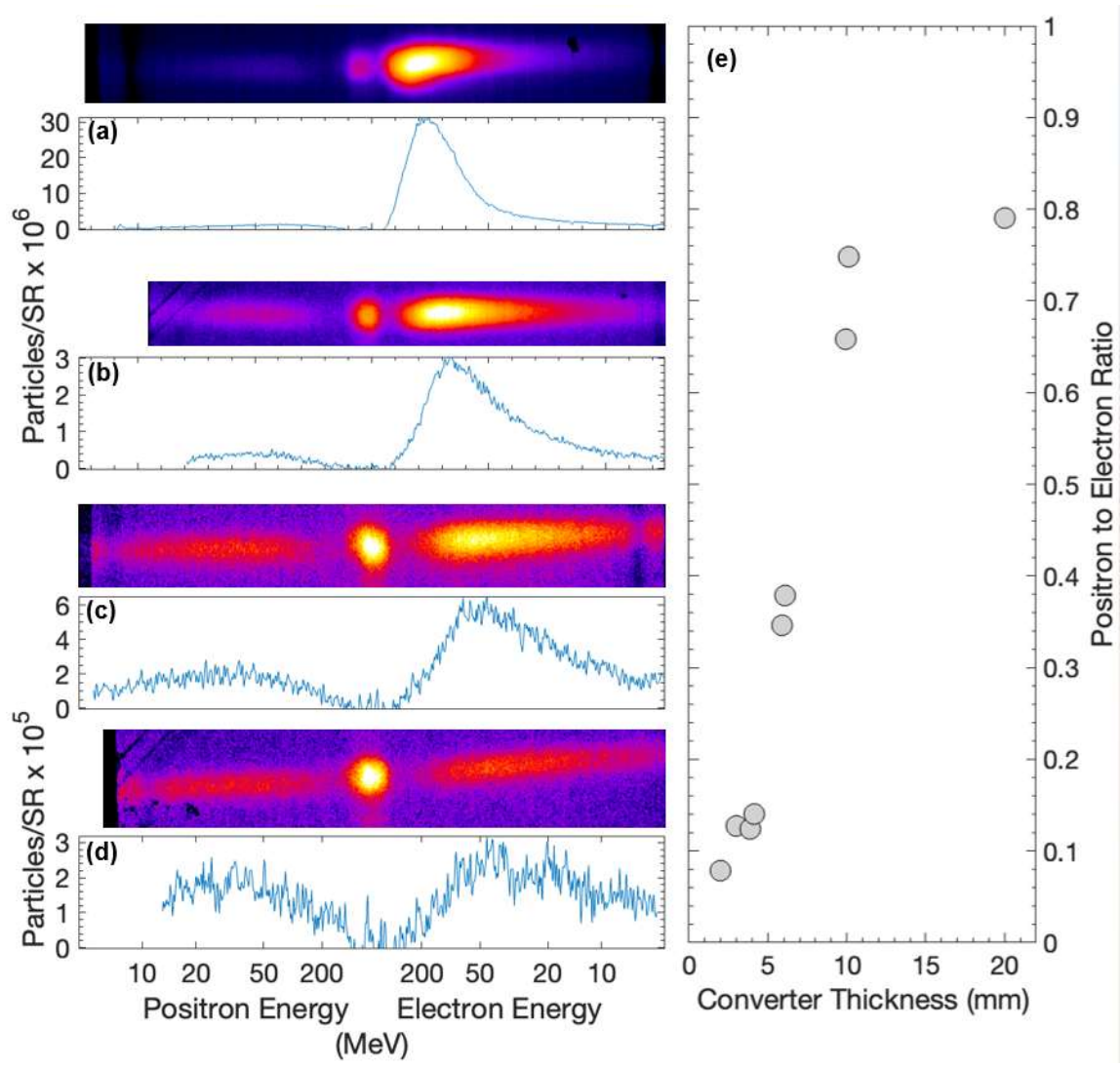

FIG 4. Measured electron-positron spectra for tungsten converter thickness of (a) 2 mm, (b) 4 mm, (c) 6 mm, (d) 10 mm. The images show the recorded image plate data, and the traces show the integrated signal as a function of energy. The magnetic field was oriented so that positrons are deflected along the negative direction on the horizontal axis and electrons in the positive direction. (e) Measured ratio of positron to electrons, each integrated in energy and angle, as a function of W converter thickness from 2 mm -20 mm.

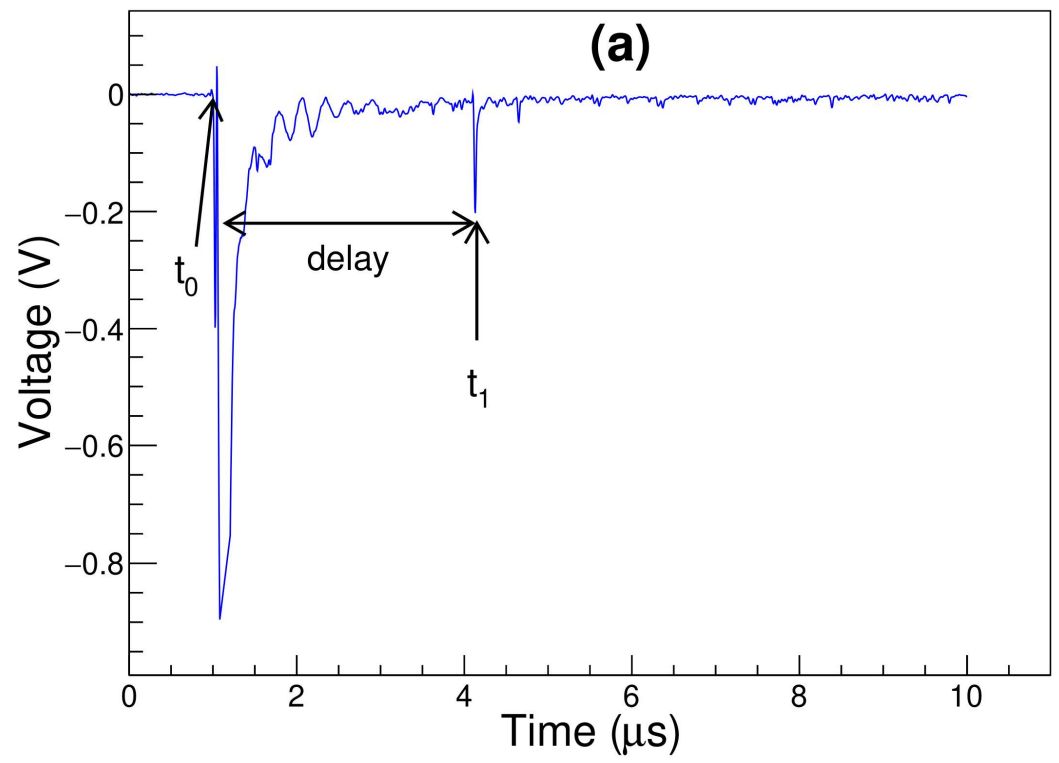

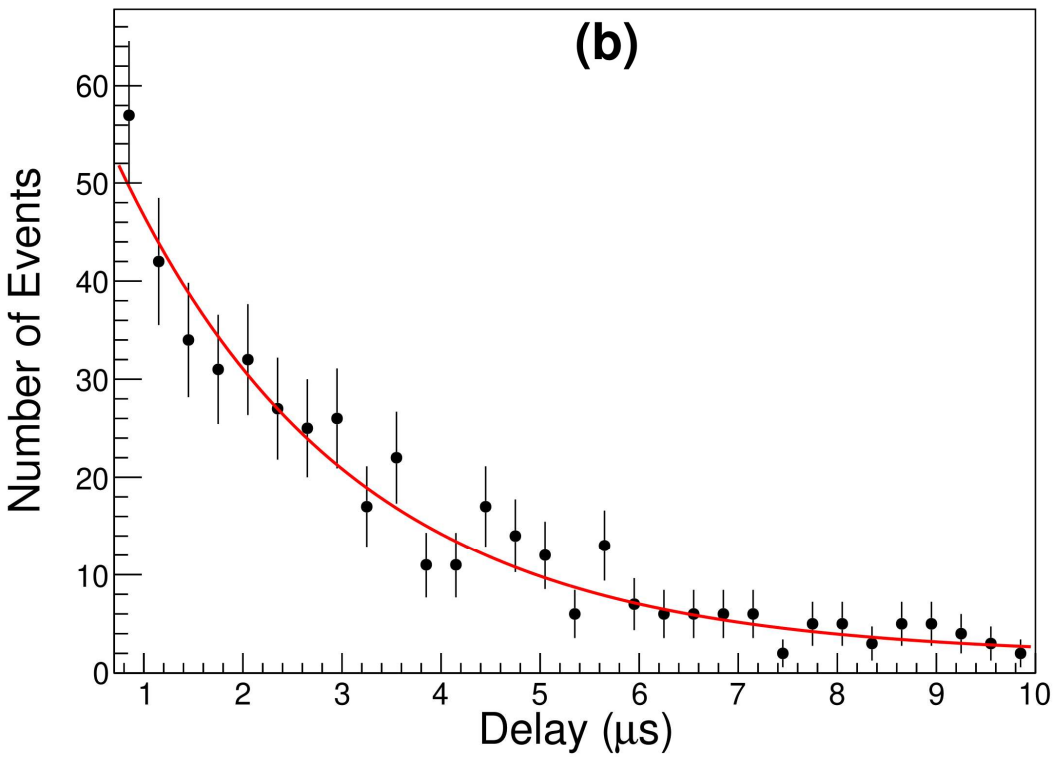

FIG 5. (a) Representative raw signal recorded by the PMT-based detector. The main saturated peak is followed by muon decay candidate peak several microseconds later. (b) Histogram showing the measured delays of 392 muon decay candidates. PMT signals recorded more than 0.7 µs after the arrival of the gamma ray spike, with amplitudes greater than 100 mV are included. The solid curve shows the least squares fit exponential decay curve with a fit decay constant. As can be seen from this data, the measured lifetime of 2.3µs is quite close to the known decay time of 2.197µs, yielding high certainty that laser-driven muons are being produced.

Measurements of particles transmitted through the lead and HDPE shielding were made using the setup shown in Fig 3(b). A representative measured raw PMT trace is shown in Fig. 5(a). In the collected data, a large impulse is observed, corresponding to the arrival of the electromagnetic shower of generated relativistic particles and gamma rays at the detector. This first pulse saturated the PMT on nearly all shots and resulted in a deadtime of approximately 0.7 µs before useful signal can again be retrieved from the detector. Following this main pulse, a series of distinct peaks are observed, some of which correspond to the decay of muons that have come to rest within the detector scintillator volume. Other processes and particles contributed to these later peaks, most notably, a high flux of neutrons that continued to arrive for many microseconds after the main pulses.

Muon candidates were identified by applying a pulse voltage threshold of 100 mV and a temporal window starting 0.7 µs to 10 µs, excluding pulses too close to the saturation of the PMT. Details on these chosen thresholds are presented in Appendix A. These candidate pulses are sorted into time delay bins with temporal width of 300ns, and a fitting function, $N_\mu(t) = N_0 e^{-t/\tau} + b$, is applied to a histogram of muon candidates, where $\tau$ is the time delay between the arrival of the speed-of-light electromagnetic shower and the muon candidate pulse, and the constant b accounts for candidate events that arise from different processes. Simulations of the detector response show that only muons produce a pure exponential decay signature on the 1–10 μs timescale. Neutron-induced signals constitute the dominant background contribution at later times. Lower amplitude peaks that may be due to neutrons are visible at the later times in the trace of Fig. 5(a). A total of 392 muon decay candidates were identified using these criteria over 4955 total laser shots recorded, including shots that did not produce electron beams. The muon decay histogram data along with the exponential fit curve are plotted in Fig. 5(b) and resulted in a lifetime of $\tau = 2.3 \pm 0.3$ μs, in agreement with the PDG value [52]. This provides strong evidence for the generation of muons by the laser-driven electron beams. Based on the highest energy and charge electron beams measured, the estimated number of muons per shot was up to $3.6 \times 10^4 - 4.7 \times 10^4$. Over the 1150 beams produced and shown in Fig. 2, an estimated 5000-6500 muons were produced per beam on average, which includes low charge and energy beams resulting in lower muon production. Details on these estimates are provided in Appendix B. Monte Carlo simulations presented below also show that a significant number of muons are generated through pion decay downstream in the beam dump and other materials.

## IV. DEMONSTRATION OF LASER-DRIVEN HIGH ENERGY PARTICLE RADIOGRAPHY

Having established the generation of electron–positron pairs and verified the production of muons through their characteristic decay signature, we performed a proof-of-principle radiography experiment to evaluate the imaging potential of the generated particle beam. The goal of this experiment was to determine whether the highly penetrating secondary particles produced by the laser wakefield accelerator could be used to detect and image dense objects after propagation through substantial shielding and building materials.

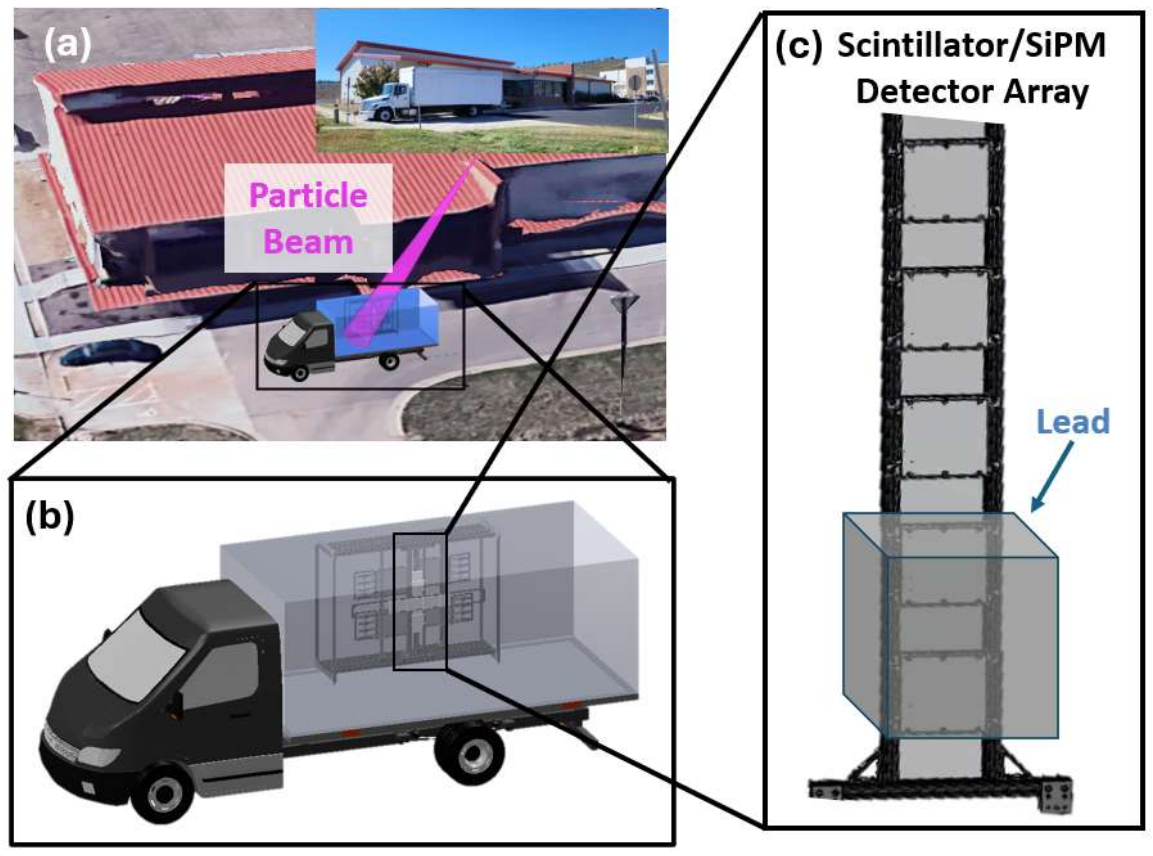

FIG 6. Conceptual diagram of the experimental setup employed for the proof-of-principle imaging demonstration. (a) Outside of the Advanced Beam Laboratory at Colorado State University a (b) truck was parked on-axis with ALEPH generated electron beam. A photo in inset. Primary electrons were stopped by the beam dump and shielding allowing primarily high energy secondary particles, including muons, to propagate outside of the building. (c) Inside the truck, an array of PMT-based detectors was placed, and an object consisting of 60 cm of lead and 15 cm HDPE could be placed to obscure two of the detectors.

The experimental configuration is shown in Fig. 6. A detector array was positioned inside a truck parked outside the laboratory building, approximately 15 m downstream of the tungsten slit assembly at the entrance to the electron spectrometer. Before reaching the detector location, the particle beam propagated through the magnetic spectrometer, the lead and HDPE shielding assembly described in Section 2, two interior laboratory walls, the exterior building wall, and the wall of the truck itself. The detector array consisted of ten EJ-200 plastic scintillator detectors coupled to silicon photomultipliers (SiPMs). The detectors were arranged in a vertical linear array aligned normal to the beam direction. The beam intersected the detector plane approximately one third of the way up from its lower edge. A test object consisting of 60 cm of lead and 15 cm of HDPE was positioned so that it could be inserted into and out of the beam.

Monte Carlo simulations using FLUKA were performed to understand the evolution of the particle beam during transport and through the object used for the radiography. The details of this simulation, which included a realistic representation of the material encountered by the beams, are in Appendix C. Figure 7(a) shows the evolution of the particle fluence along the beam path near the object and detectors. The increase in the total particle population immediately inside the object results from the generation of secondary particles in the electromagnetic shower, which is dominated by relatively low-energy particles and neutrons. As propagation continues, these secondary particles are rapidly attenuated by the shielding and surrounding materials, while the more penetrating muon component decreases much more slowly.

The corresponding evolution of the beam composition is shown in Fig. 7(b). Near the source, the particle population contains a substantial contribution from neutrons and other secondary particles. After transmission through the object and subsequent propagation, however, the surviving beam becomes increasingly enriched in muons.

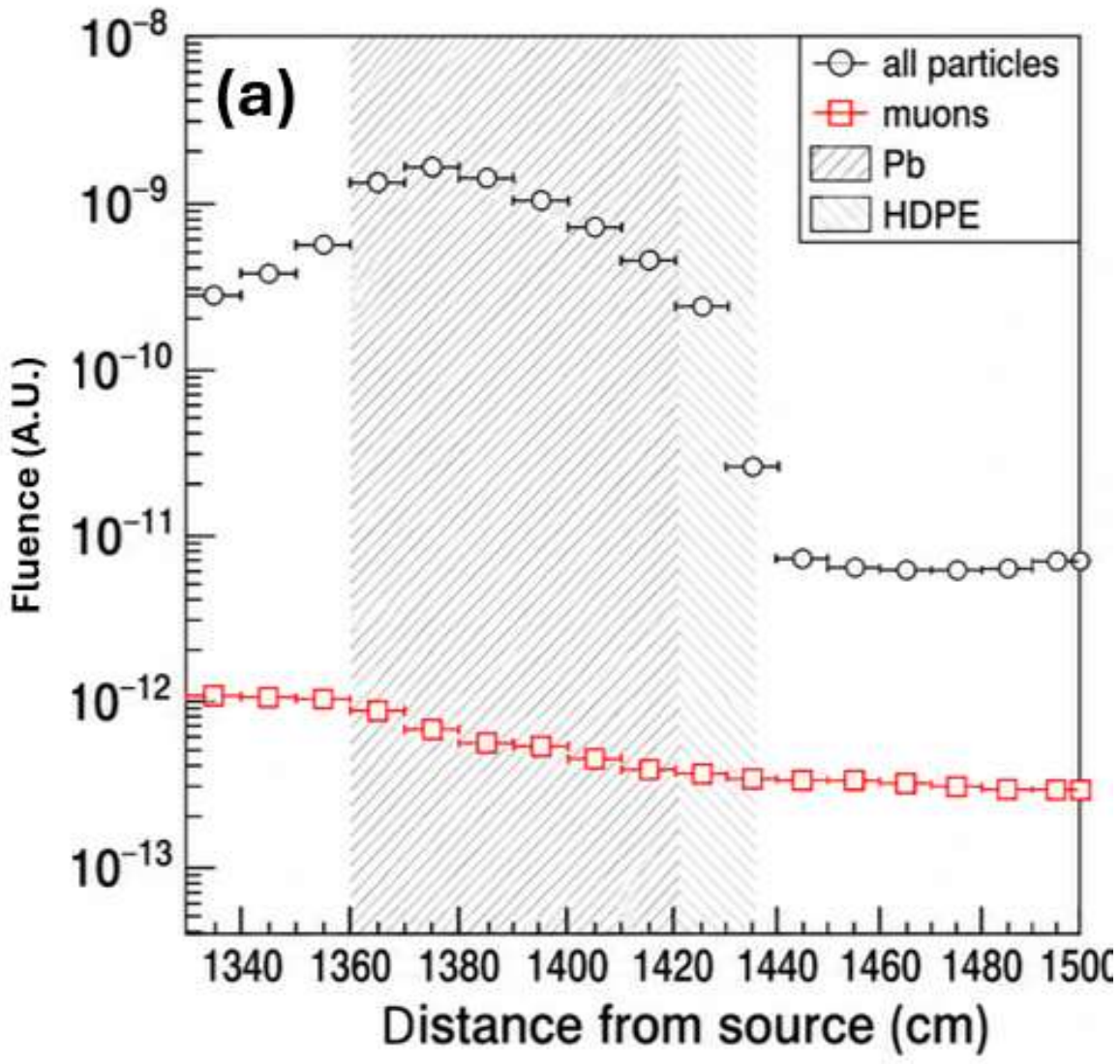

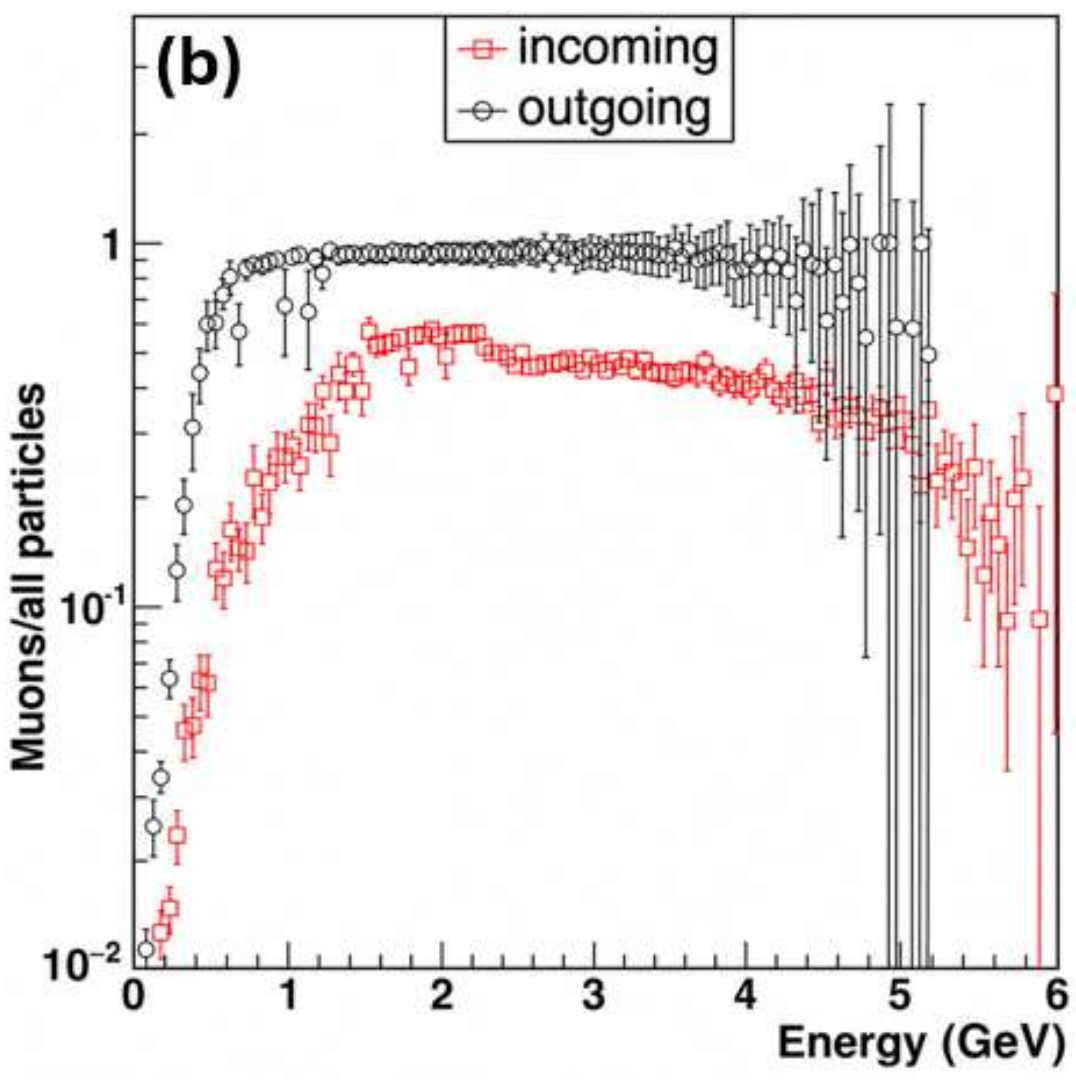

FIG 7. Results of the Monte Carlo simulation of the particle flux at the imaging object. (a) Simulated flux of all particles and only muons traversing the 60cm lead and 15cm HDPE object. The all particle density increases upon entry into the lead object because of the generation of additional secondary particles (b) Simulated ratio of muons to all particles immediately before and after the lead object as a function of energy. As can be seen from these plots, the lead attenuation of all particles results in a decrease of nearly two orders of magnitude, while resulting in only small reduction in the overall flux of muons. The particles exiting the object with energies >1 GeV is nearly entirely muons.

Simulations of the transmitted particle spectrum show that neutrons dominate the particle population below approximately 600 MeV, whereas above this energy muons account for approximately 90% of the transmitted particles. Consequently, over the energy range most relevant for penetration through dense materials, the transmitted beam is strongly muon-rich, supporting its suitability for radiographic applications.

Despite propagation through the spectrometer, shielding, building materials, and the ~15 m transport distance, a readily detectable particle flux was observed on every LWFA shot. Although the detector system does not provide particle identification, simulations indicate that the detected signal arises from a mixture of muons, neutrons, pions, and secondary radiation generated by their interactions with surrounding materials.

To demonstrate imaging capability, two data sets were acquired. In the first configuration, the detector array was exposed directly to the transmitted particle beam. In the second configuration, a 60 cm thick lead and 15 cm HDPE absorber was positioned in front of two detectors (labeled #7 and #8), partially obscuring them from the direct beam. A total of 82 laser shots were recorded without the object in the beam and 75 shots were recorded with the object in place.

The presence of the absorber object produced a pronounced reduction in signal amplitude in the obscured detectors while leaving the remaining detectors largely unaffected. This behavior is evident in the signal distributions shown in the histogram of Fig. 8. Figure 9 presents the measured detector responses for representative single shots and for the average over the full data set. In both cases, the shadow cast by the lead absorber is clearly visible on the corresponding detectors. Monte Carlo particle simulations of the setup, presented in Appendix C, and taking into account the expected relative detector sensitivities show that the contribution to the observed change in signal due to the shadow on the highest charge beams is estimated to be approximately 20%, 45%, and 30%, for muons, neutrons, and pions, respectively. The remaining ~5% is due to small quantities of other particles. This LWFA-driven multi-species shadow, produced by the combination of attenuation of muons, pions, and neutrons, and dominated by the most penetrating muon at high energies, distinguishes this technique as a useful compact source for high energy particle radiography.

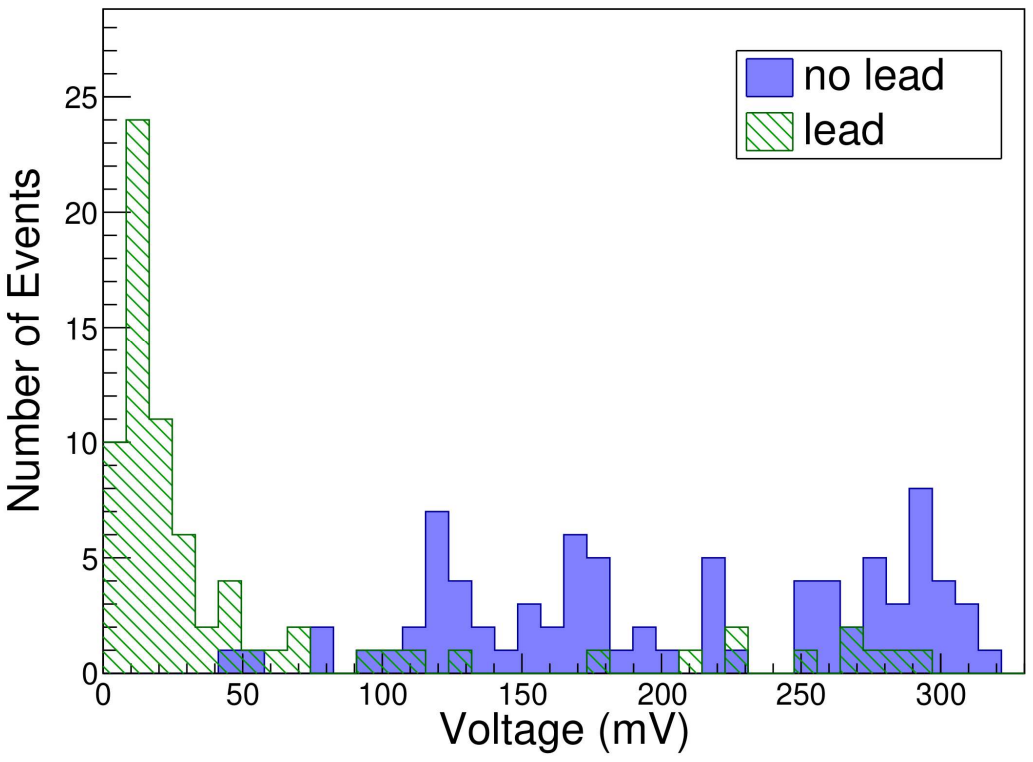

FIG 8. (a) Amplitude distribution of an obscured detector. With object in place, the distribution becomes substantially narrower and shifts toward lower amplitudes, indicating strong attenuation of the particle flux. In contrast, the unobscured detector (b) continues to exhibit a broad distribution, reflecting persistent shot-to-shot fluctuations in beam intensity.

The ability to resolve the object on individual laser shots is particularly significant. Conventional muography relies on naturally occurring cosmic-ray muons and often requires acquisition times ranging from days to months to obtain a useful image. In contrast, the present results demonstrate the potential for single-shot particle radiography using a compact laser-driven source. Our simulations show that on the highest energy and charge beams generated, the muon flux incident on the object is ~50 muons/cm$^2$. This single-shot flux is equivalent to >8 hours of integration

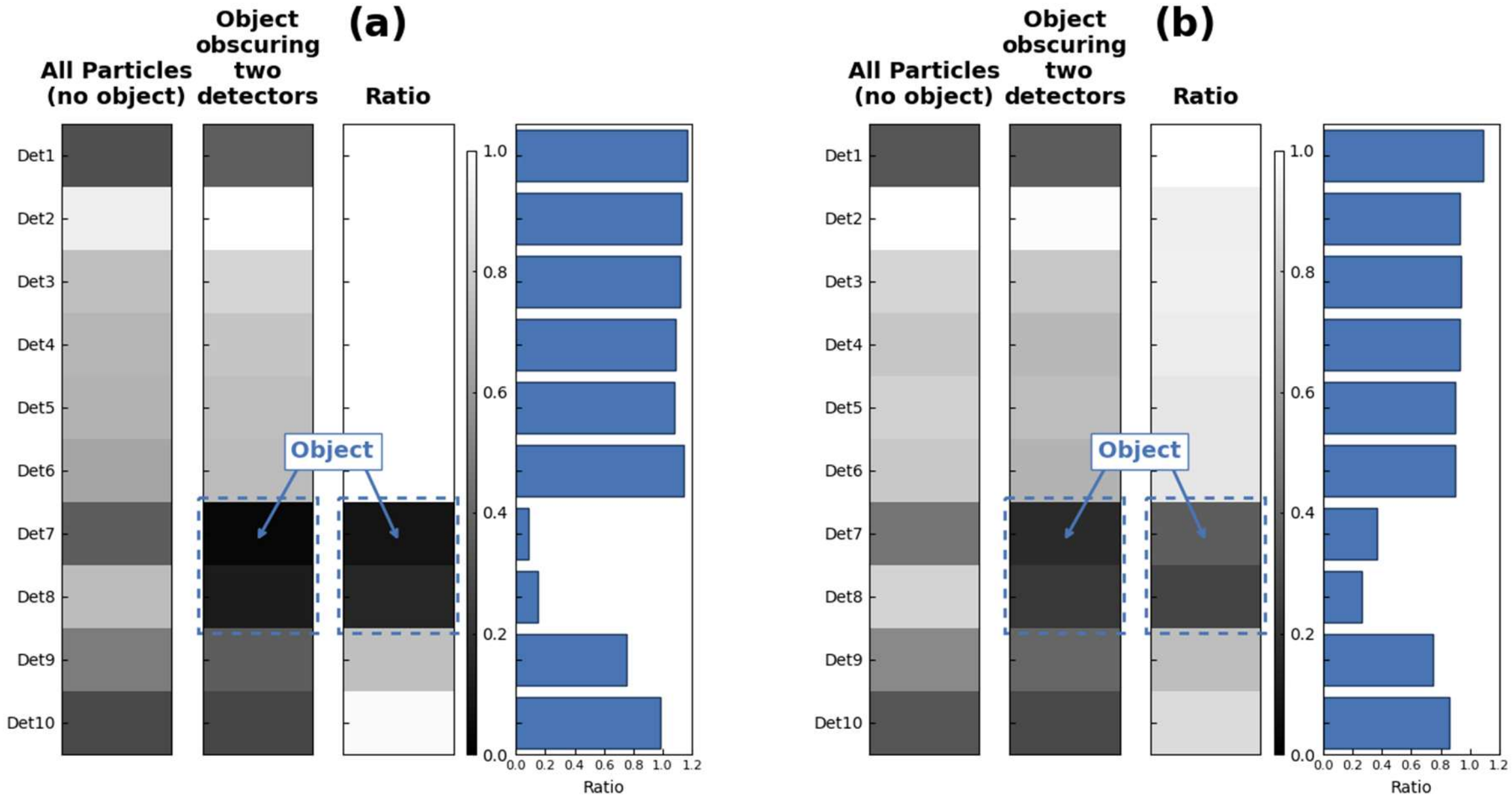

FIG 9. Summary of proof-of-principle radiography result. Detector array signals recorded with and without a 60 cm lead object placed in front of two of the detectors for (a) a single shot of the LWFA; (b) the mean results over 82 no-lead shots and 75 shots with a 60 cm lead object. As can be seen from these images, despite the particle beam travelling through multiple interior and exterior walls, lead shielding, the presence of material in the beamline is clearly observed, even on a single shot

of naturally occurring muons near the horizon. Although the detected signal contains contributions from neutrons and pions in addition to muons, this proof-of-principle experiment establishes that the secondary particle beam generated by a multi-GeV laser wakefield accelerator possesses sufficient penetration and flux to image dense objects and structures through substantial shielding.

## V. SUMMARY

We made a first demonstration of radiography using a multi-GeV LWFA-generated particle beam consisting of muons, pions, and neutrons. Despite propagation through a thick lead beam dump and shielding, multiple building walls and a truck wall, and approximately 15 m of transport, a readily detectable particle flux was observed outside the laboratory. A significant component of the beam was muons, verified through measurement of their characteristic decay signature in a large-volume scintillator detector. The measured decay lifetime, $\tau = 2.3 \pm 0.3$ μs, is in good agreement with the PDG value, providing strong evidence for muon production by the laser-driven electron beam. In parallel, the generation of electron–positron pairs through the Bethe–Heitler process was characterized using a dedicated magnetic spectrometer. The observed evolution of the electron and positron spectra with converter thickness is consistent with the development of an electromagnetic cascade and provides an experimental benchmark of secondary-particle production in the target.

A dense lead object was successfully imaged by a detector array positioned inside a truck parked outside the building, and the object shadow was clearly resolved on individual laser shots. Simulations showed that the high energy particles transmitted through the object consisted almost entirely of muons, demonstrating their deep penetration ability for radiography of large dense structures. Despite transmission through the beam dump, shielding, and the walls of the building, the inferred muon flux on a single shot on the large detector array is equivalent to >8 hours of integration of the natural, cosmic-ray-generated flux of muons near the horizon. This result demonstrates the potential of compact laser-driven sources for single-shot interrogation of dense structures, in contrast to conventional muography techniques that rely on long integration times using naturally occurring cosmic-ray muons. For imaging applications at large distances and through dense materials, we expect that our current LWFA-driven source is capable of single-shot imaging dominated by muons.

These results establish that multi-GeV laser wakefield accelerators can serve as compact drivers of highly penetrating secondary particle beams containing muons and provide a pathway toward transportable radiographic systems capable of probing objects and structures inaccessible to conventional imaging methods.

Future work will focus on improved characterization of the generated particle flux, spectra, and spatial distributions, as well as the development of optimized detector systems for radiographic and tomographic applications. Continued advances in laser wakefield acceleration are expected to increase both particle energy and flux, further extending the range of potential imaging applications. As recently reported, single stage, laser-driven accelerators capable of achieving up to 100 GeV electron bunches are feasible drivers of high flux, high energy muon sources [45]. These compact accelerators, powered by next generation efficient and compact laser drivers [62] based on bulk Tm:YLF [62,63], coherently combined ultrafast fiber lasers [64–66], or cryogenically-cooled Yb:YAG [67,68], could operate at kHz repetition rates. Assuming the same electron beams and geometry in the demonstrated experiment, these systems would result in a muon source with 30 million times higher flux than horizontal cosmic sources, enabling radiography of dense structures impenetrable to any other known beams.

## ACKNOWLEDGEMENTS

This work was supported by Defense Advanced Research Projects Agency (DARPA) under the Muons for Science and Security (MuS2) Program. The experiments were conducted with the ALEPH laser at Colorado State University's Advanced Beam Laboratory funded by LaserNetUS under U.S. Department of Energy Award DE-SC0021246. The University of Maryland team acknowledges support from the U.S. Department of Energy (DE-SC0015516 and DE-SC0024398). Jorge J. Rocca acknowledges support from a Vannevar Bush Faculty Fellowship, Office of Naval Research Award 000142012842. This work was performed under the auspices of the U.S. Department of Energy by Lawrence Livermore National Laboratory under contract DE-AC52-07NA27344. LLNL-JRNL-2021394. The authors thank John Liptac and Mark Wrobel for early conversations that greatly contributed to this work.

## DATA AVAILABILITY

The data that support the findings of this article are not publicly available. The data are available from the authors upon reasonable request.

## APPENDIX A: MUON DETECTOR CONSTRUCTION AND VERIFICATION

### 1. Muon Detector Geometry

The configuration of the muon lifetime measurement consisted of a high-sensitivity EJ-200 plastic scintillator block optically coupled to a 125 mm diameter Hamamatsu R1250 photomultiplier tube (PMT). The structural details and physical dimensions are illustrated in the computer-aided design (CAD) model in Fig. 10(a). The active volume of the plastic scintillator was 75mm x 150 mm x 300mm. To reduce the contribution of other secondary particles, such as gamma rays and neutrons, the entire detector was enclosed in a light-tight housing, which was further encapsulated within a robust multi-layer passive shielding tracking layout. This shielding assembly surrounded the core components with 100 mm of lead and at least 50 mm of high-density polyethylene (HDPE) on all sides.

A 3D CAD model of the detector units used in the radiography demonstration is shown in Fig. 10(b). A 150 mm x 150 mm x 25 mm block of EJ-200 plastic scintillator was wrapped in vikuiti film on five of its faces to maximize signal detection. Two 6 mm silicon photomultipliers were mechanically attached to the remaining face with index-matching grease. The SiPMs were biased and the detected signals coincident with firing the laser were digitized by a CAEN 2740 digitizer with 20 MHz analog bandwidth and 125 MS/s sampling rate. 10 detectors were arrayed vertically in a truck parked in line with the beam outside of the laboratory building. Although the detectors used in the experiment were not calibrated and cannot provide discrimination between particle types, it is expected that they are 5-10x more sensitive per particle to muons and pions than to neutrons for the energies encountered here [52,69].

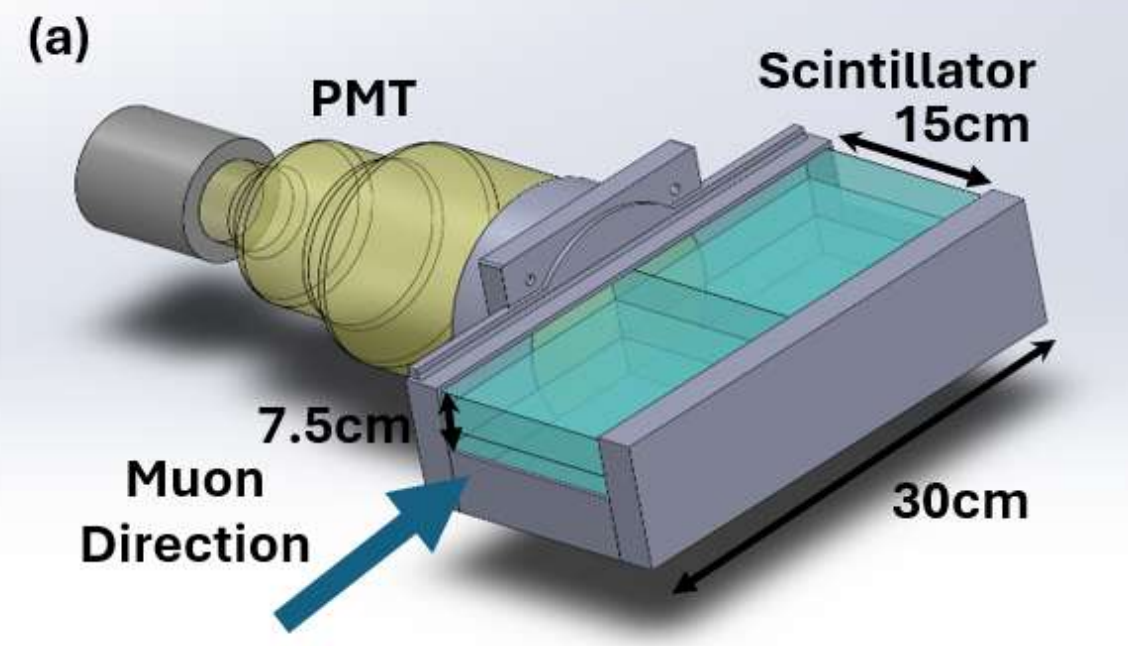

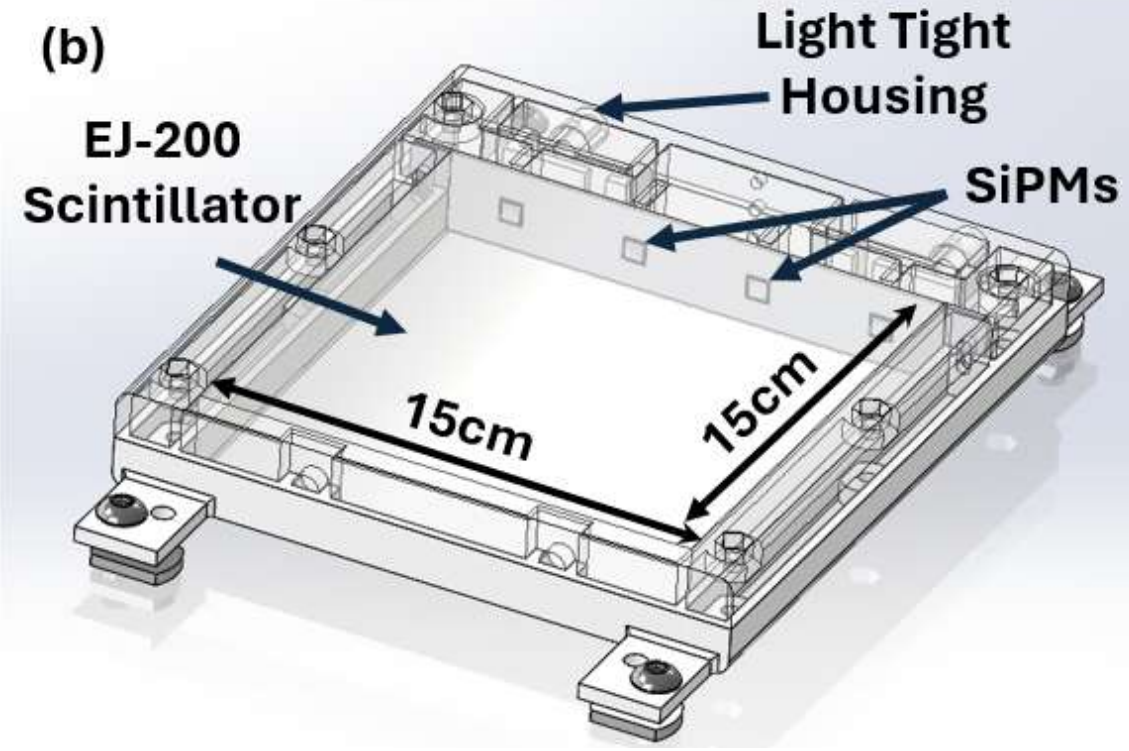

FIG 10. Three-dimensional CAD renderings of (a) the muon decay detector and (b) the detectors used in the radiography demonstration.

### 2. Calibration Using Cosmic Ray Muons

A cosmic ray muon calibration run was conducted to validate the detector's response and the reliability of the fitting framework. To maintain systematic consistency with the electron beam analysis, the same lower cutoff time of 0.7μs was identically applied to the cosmic ray decay fit. With this analysis window, the calibration data were analyzed under two amplitude threshold scenarios to verify the stability of the signal extraction. These results are plotted in Fig. 11(a) and (b). Applying the 100mV threshold, identical to the laser generated beam analysis, this yielded 136 muon candidates with an extracted lifetime of $2.12 \pm 0.49$ μs. Conversely, when no threshold was applied, the acceptance was maximized to capture 1208 muon candidates, resulting in an extracted lifetime of $2.13 \pm 0.12$ μs. Both of these are consistent with the (PDG) value of $2.1969811 \pm 0.0000022$ μs within statistical uncertainty. This indicates that the choice of the amplitude threshold does not significantly bias the determined decay times.

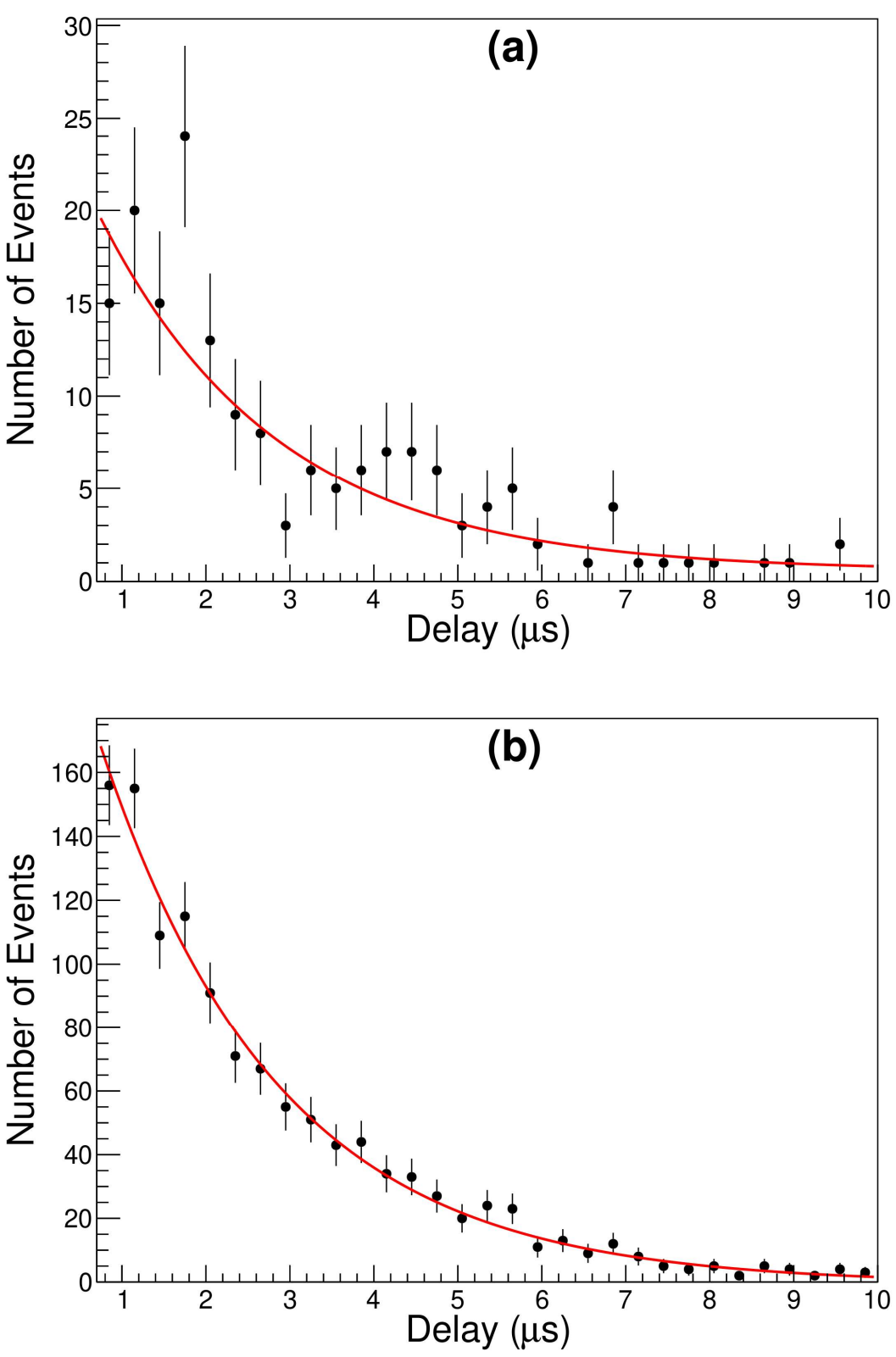

FIG 11. Cosmic ray muon decay time distributions and the corresponding exponential fits under two different amplitude threshold configurations: (a) with the 100mV threshold applied (yielding 136 pure muon candidates with a fitted lifetime of 2.12 ± 0.49 μs) and (b) with no amplitude threshold applied to maximize acceptance (capturing 1208 muon candidates with a fitted lifetime of 2.13 ± 0.12 μs). Both histograms are fitted from a unified lower cutoff time of 0.7μs up to 10.0μs.

### 3. Discrimination Parameter Scans and Stability Study

To evaluate the systematic uncertainties introduced by the choice of the cutoff time and the amplitude threshold, comprehensive discrimination scans were performed across both variables independently. Specifically, the amplitude threshold scan was conducted at a fixed cutoff time of 0.7μs while the cutoff time scan was carried out at the amplitude threshold of 100mV. The fitted lifetime results as a function of these two parameters are presented in Fig. 12(a) and Fig. 12(b), respectively.

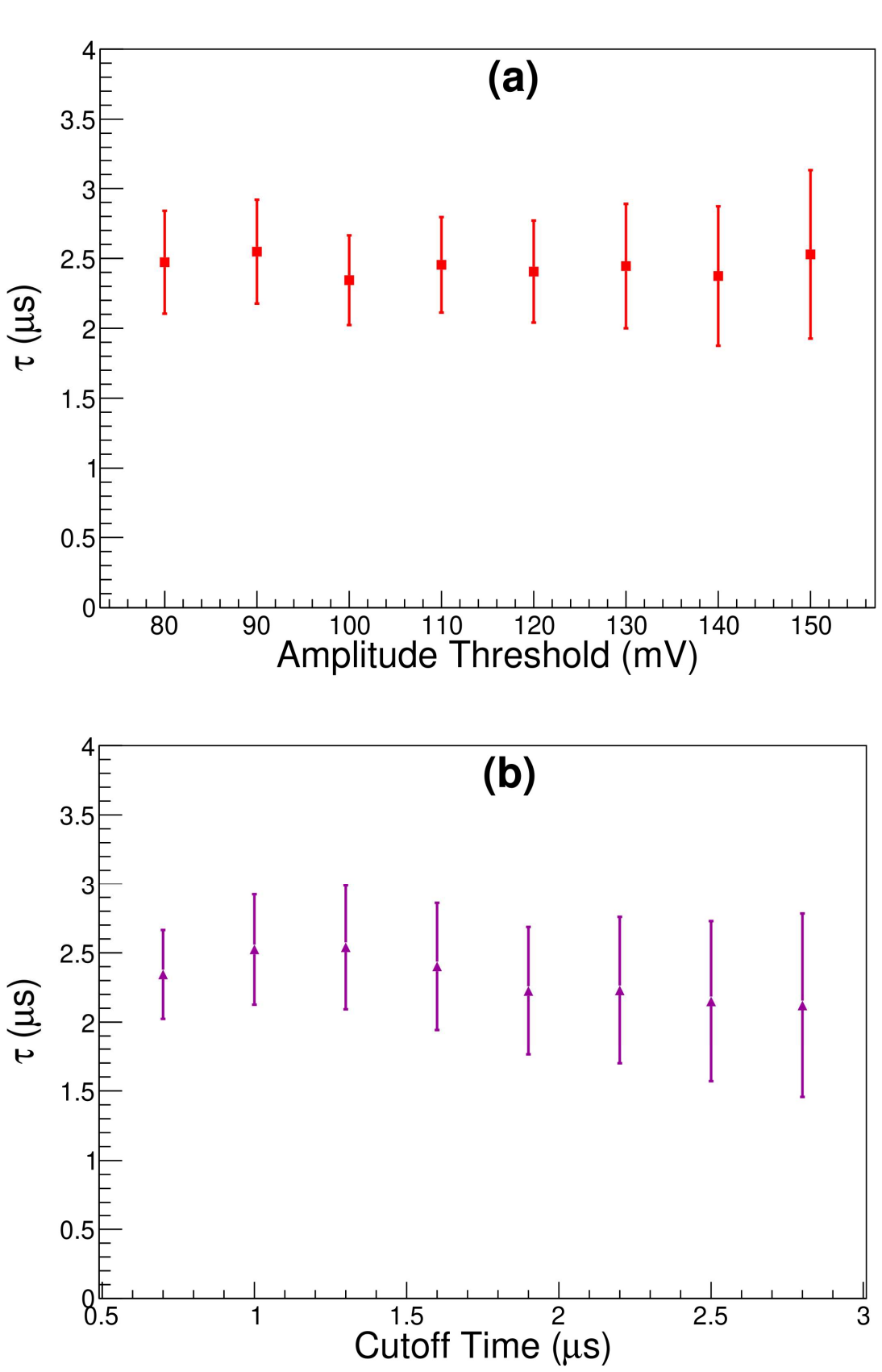

FIG 12. Fitted muon lifetime as a function of (a) the amplitude discrimination threshold with a fixed cutoff time of 0.7μs, and (b) the cutoff time with a fixed amplitude threshold of 100mV.

As illustrated in Fig. 12(a), the fitted muon lifetime remains stable across the entire scanned range of amplitude thresholds from 80 mV to 150 mV, even as the number of muon candidates decreases from 559 down to 141. Similarly, the cutoff time scan shown in Fig. 12(b) exhibits no drift as the fit start time is progressively advanced from 0.7μs to 2.8μs, with the corresponding muon candi-dates dropping from 392 to 162. Consequently, the chosen configuration of a 0.7μs cutoff time and a 100 mV threshold represents a secure and reliable phase space for the primary physics analysis.

## APPENDIX B: MUON YIELD ESTIMATES

An estimate of the maximum LWFA-driven muons produced was made using three methods using the measured electron spectrum shown in Fig. 13(a). First, we used a Geant4 Monte Carlo particle transport model to simulate muon production efficiency as a function of primary electron energy. In this simulation, monochromatic, collimated beams of electrons of varying energies were projected into a 2 cm thick tungsten converter target, and the muons produced through Bethe-Heitler pair production exiting the converter were counted. Our simulations have shown that once a converter target of sufficient thickness is employed, the efficiency of conversion to muons is relatively insensitive to further increases in converter thickness. Figure 13(b) displays the results of this simulation, which shows that 1-10 GeV electron beams produce 1-100 muons per pC charge, respectively. These simulated conversion efficiencies are in good agreement with other simulations of BH pair production of muons recently reported [28]. Applying this conversion efficiency to the measured electron spectrum of Fig. 13(a), which contains ~1.4nC total charge centered near 5 GeV yields a total of 4.0 x $10^4$ muons. Second, interpolating the equivalent conversion efficiency reported in [28] and applying to the electron beam spectrum yields a total 3.6x$10^4$ muons, which is in close agreement with our simulation. Finally, we used the FLUKA particle simulation code described in Appendix C to estimate the yield of muons to be 4.7x$10^4$ by integrating the flux in a plane inside the lead beam dump by which the vast majority of primary electrons have been stopped. This number is higher than the two other methods because muons generated through pion decay were also included. The contribution of pion decay generated muons will be significant for most shielded imaging applications.

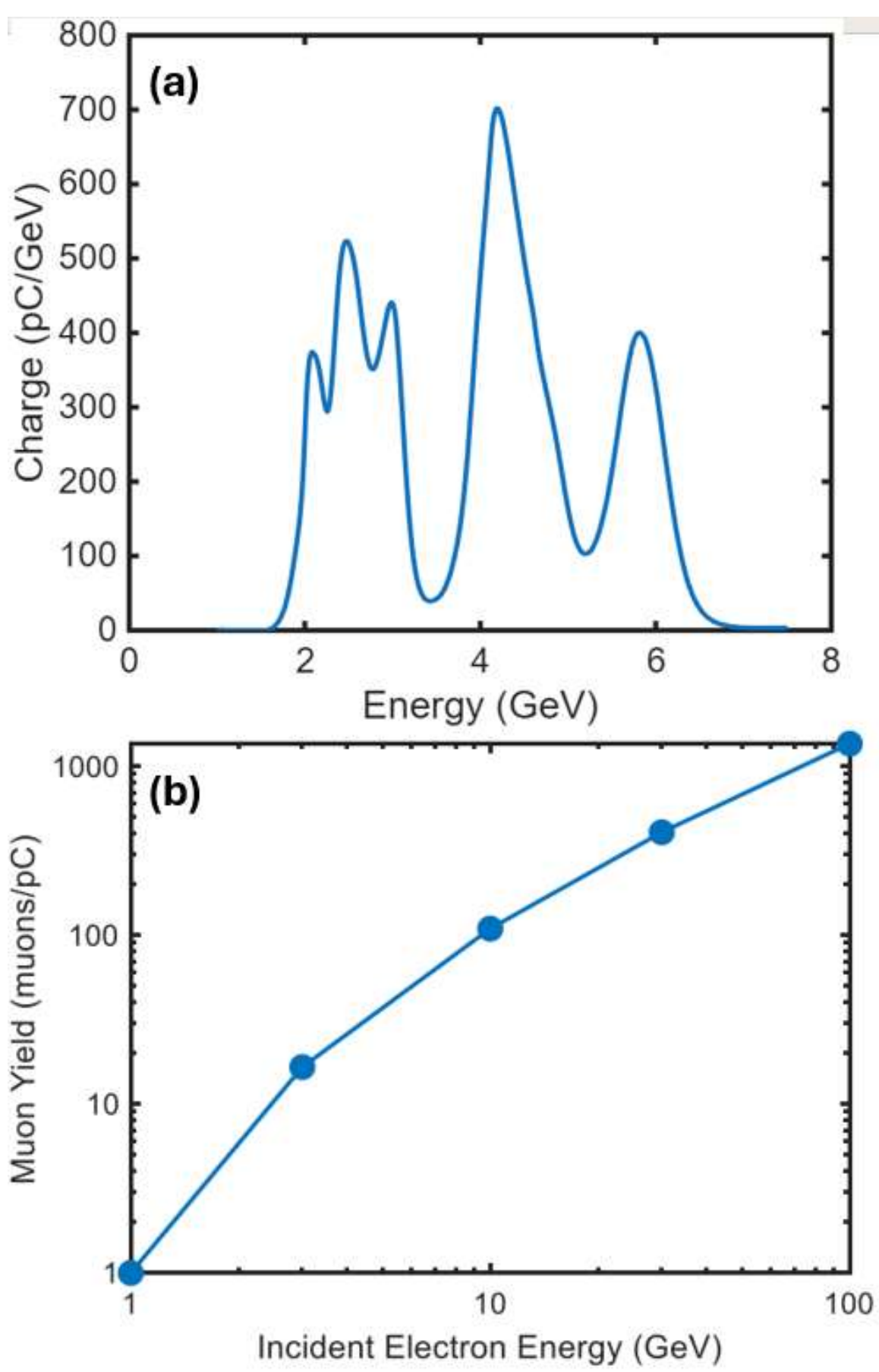

FIG 13. (a) Measured spectrum of a high charge and high energy LWFA electron beam. (b) Simulated conversion efficiency of GeV electrons to muons through BH pair production.

## APPENDIX C: MONTE CARLO PARTICLE RADIOGRAPHY SIMULATIONS

FLUKA Monte Carlo particle transport and interaction simulations were employed to interpret the high energy particle radiography demonstration experiments. For these simulations, the source term was the measured electron beam spectrum shown in Fig. 13(a). A total number of $8 \times 10^{12}$ primary electrons (equivalent to many beams) were simulated. Transport thresholds for the electromagnetic part were set at 10 MeV, leading particle biasing was used in heavy materials (lead, tungsten, steel). Relevant physics processes, such as BH pair production, bremsstrahlung, photo-pion and photo-neutron production were all included in the analysis of the secondaries. A realistic representation of the relevant materials and geometry was used, that included the magnetic electron spectrometer collimator, magnet materials and field, and housing, detector shielding, and beam dump as well as the ~15 m propagation to the detector array. Figure 14(a) and (b) show the resulting simulated particle spectra immediately before and after the lead and HDPE target. As can be seen from these figures, at energies below 600 MeV neutrons were the dominant species before and after the object, with muons and pions dominating at higher energies. At energies >100 MeV, the flux of muons at the object was simulated to be ~50 $cm^{-2}$, with a nearly identical flux of pions and approximately 60x higher flux of neutrons. On transmission through the object, the muons were attenuated by a factor of ~3x, neutrons by ~20x, and pions by ~40x. Considering the relative sensitivities of the detectors, this results in significant contributions to the observed shadow from all three species in the experiment.

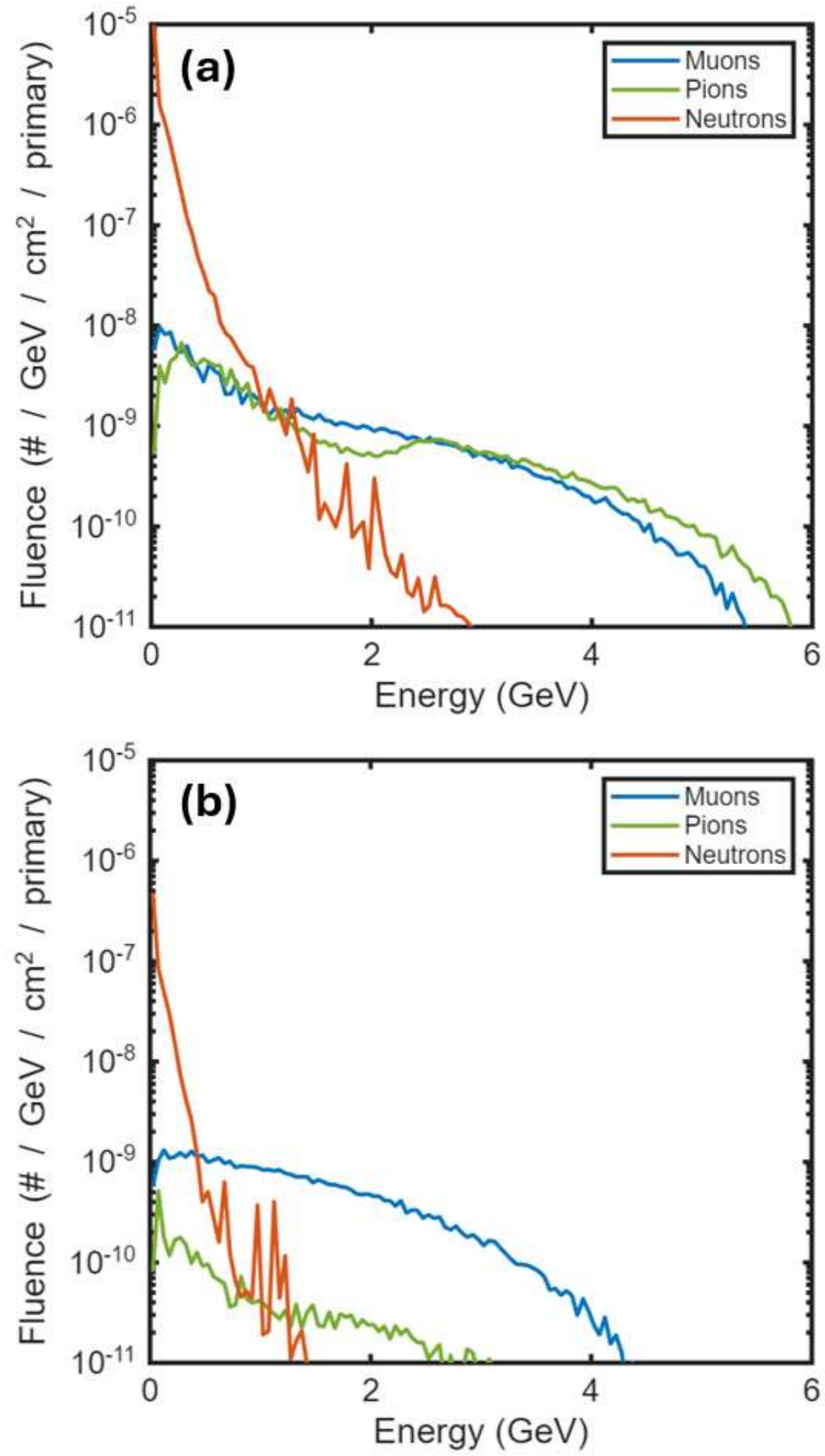

FIG 14. (a) Simulated spectra of muons, pions, and neutrons incident on the object used in the radiography experimental demonstration. (b) Simulated spectra of the particles exiting the object.